# A Novel Clustering-Based Algorithm for Continuous and Non-invasive Cuff-Less Blood Pressure Estimation

Ali Farki,[1] Reza Baradaran Kazemzadeh,[1] and Elham Akhondzadeh Noughabi [1]

[1] Department of Information Technology Engineering, Industrial and Systems Engineering Faculty, Tarbiat Modares University, Tehran, Iran

Correspondence should be addressed to Ali Farki; Alifarki@modares.ac.ir

## Abstract

Extensive research has been performed on continuous, non-invasive, cuffless blood pressure (BP) measurement using artificial intelligence algorithms. This approach involves extracting certain features from physiological signals like ECG, PPG, ICG, BCG, etc. as independent variables and extracting features from Arterial Blood Pressure (ABP) signals as dependent variables, and then using machine learning algorithms to develop a blood pressure estimation model based on these data. The greatest challenge of this field is the insufficient accuracy of estimation models. This paper proposes a novel blood pressure estimation method with a clustering step for accuracy improvement. The proposed method involves extracting Pulse Transit Time (PTT), PPG Intensity Ratio (PIR), and Heart Rate (HR) features from Electrocardiogram (ECG) and Photoplethysmogram (PPG) signals as the inputs of clustering and regression, extracting Systolic Blood Pressure (SBP) and Diastolic Blood Pressure (DBP) features from ABP signals as dependent variables, and finally developing regression models by applying Gradient Boosting Regression (GBR), Random Forest Regression (RFR), and Multilayer Perceptron Regression (MLP) on each cluster. The method was implemented using the MIMICII dataset with the silhouette criterion used to determine the optimal number of clusters. The results showed that because of the inconsistency, high dispersion, and multi-trend behavior of the extracted features vectors, the accuracy can be significantly improved by running a clustering algorithm and then developing a regression model on each cluster, and finally weighted averaging of the results based on the error of each cluster. When implemented with 5 clusters and GBR, this approach yielded an MAE of 2.56 for SBP estimates and 2.23 for DBP estimates, which were significantly better than the best results without clustering (DBP: 6.27, SBP: 6.36).

## 1.Introduction

Blood pressure (BP) is one of the most important health indicators and can be used to diagnose various diseases. BP measurement techniques can be broken down into two categories of Invasive methods and Non-invasive methods. While the invasive approach tends to provide more accurate BP readings, it has some drawbacks and limitations. The World Health

Organization has issued reports on the subject that each year, 9.4 million people die from excessive blood pressure around the world (hypertension) , and roughly 30% of all men and 25% of all women suffer from this condition[1]. After diabetes, hypertension is the second leading cause of cardiovascular disease, but it also tends to be asymptomatic, so it has been called the silent killer. As one of the vital signs, blood pressure needs to be regularly controlled. In many clinical settings, BP monitoring needs to be constant , especially if the patient is old or is in the intensive care unit (ICU). Regular BP monitoring can also help prevent stroke, heart attack, and heart failure[2]–[4]. Unfortunately, most people with hypertension are unaware of their condition and how it harms their internal organs like the brain, eyes, and kidneys over time.

As mentioned earlier, there are two types of blood pressure measurement methods: Invasive and Non-Invasive. In Invasive Blood Pressure (IBP) monitoring, measurements are done by a sensor or cannula needle inserted in a blood vessel. This method can provide continuous accurate BP information but has drawbacks such as vessel blockage and potential area infection[5]. Non-invasive Blood Pressure (NIBP) monitoring methods can be classified into two categories: 1- the auditory methods and 2- the methods based on vital signals. The auditory method is the common BP measurement method which involves wrapping a cuff around the arm. Naturally, this method measures the blood pressure at one instant and cannot provide continuous BP readings. Also, using this method multiple consecutively leads to patient dissatisfaction[6]. Given the limitations of direct BP measurement methods, several indirect methods have also been developed for this purpose. As of this writing, researchers have not found a consistent relationship between Blood Pressure and Electrocardiogram and Photoplethysmogram signals ,so that blood pressure can not be reliably obtained from these signals. However, there are indeed some relationships between blood pressure and the features extracted from these signals[7], [8]. Therefore, these features can be used to create prediction models for BP estimates using data analysis methods and technologies. In the non-invasive and cuff-less BP estimation method, we first extract a vector of physiological features from ECG and PPG signals and then develop a regression model for BP estimation with these features used as input[9], [10]. The greatest weakness of non-invasive cuff-less methods compared to other BP measurement methods is their lower accuracy, which can be somewhat improved by using a combination of different features and different machine learning and data mining methods.

Over the years, researchers have conducted many studies on feature extraction from physiological signals such as ECG, PPG, ICG, and BCG and also blood pressure (BP) estimation based on these features. As mentioned, the main challenge in this field is how to raise the accuracy of BP estimates. In this paper, we introduce a new clustering-based method to achieve significant accuracy improvement in this area. This method starts with extracting PTT, PIR, and HR features from ECG and PPG signals and extracting SBP and DBP features from the corresponding ABP signal. While previous methods of this field follow this step by developing a model based on the extracted features, In all the other works, no attention has been paid to the high dispersion of data that extracted from ECG,PPG and ABP signals , which will have a negative effect on the accuracy of the model. in the proposed method, first, a clustering algorithm is applied to PTT, PIR, and HR, and then a model is developed separately for each resulting cluster using the corresponding SBP and DBP data. Since the data of the extracted features tend to have high dispersion and contain multiple trends, using the clustering algorithm in this way can greatly improve the accuracy of estimations. In many works, such as [11] and [12] , a large number of features are extracted from the raw ECG and PPG signals. According to research, by increasing the number of effective features in the development of the machine learning model, the accuracy of the model can be significantly increased. On the other hand, it can be concluded that increasing the number of extracted features can lead to high computational complexity in real-world applications. However, in our work, only 3 features have been extracted from ECG and PPG signals. Finally, the accuracy has been improved by using the clustering algorithm.

Another noteworthy point is that in the various studies that used the MIMIC dataset as their database , the researchers had no idea about the patient's physiological condition, But in our work, after extracting the features and clustering, we noticed similarities in the raw ECG, PPG and ABP signals corresponding to the data samples in each cluster. Which can be used to patients clustering, which can have a positive effect on the accuracy or correctness of features. In other works, which uses ECG and PPG signals, the appearance of each person's signal can be different, which will affect the accuracy of the extraction features and feature extraction algorithms [9], [13]. The extraction process can be more accurate by using the clustering technique and clustering the raw signals of patients based on their similarities.

## 2. Materials and Methods

*2.1 Dataset.* The MIMIC-II (Multiparameter Intelligent Monitoring in Intensive Care) dataset from the Physionet website was used in this research. This dataset contains 12,000 records of vital signals captured from people admitted to American medical centers and hospitals. The signals of this dataset include ECG, PPG, and Arterial Blood Pressure (ABP) at a sampling rate of 125Hz[14]. A preprocessed and cleaned version of this dataset is publically available on the Kaggle website[15].

*2.2. Features extraction.* The Pulse Transit Time (PTT), which is the time it takes for the arterial pulse wave to move from the aortic valve to the peripheral artery, is a typical approach to make continuous BP measurements. In other words, The time difference between the R-peak of the ECG signal and a reference point on the PPG signal of the corresponding pulse wave is referred to as the PTT[14]. The heart of this strategy is the notion of pulse wave velocity (PWV), which is obtained from the Moens-Korteweg equation (MK)[16]:

$$PWV = \sqrt{\frac{Eh}{d}} \qquad (1)$$

In this equation, E is the elastic modulus of the arterial wall, h is the thickness of the wall, ρ is blood density, and d is the vessel radius. The following formula shows how PWV is inversely correlated to PTT[17]:

$$PWV = \frac{K}{PTT} \qquad (2)$$

The distance between the heart and the reference peripheral (e.g., the fingertip) site is denoted by the K. The use of PWV leads to obtaining a more accurate PTT but requires parameters such as the person's physical characteristics[2], [15], [18], [19].

The ECG and PPG can derive the PTT features by taking the second derivative of the PPG or SDPPG signal. PTT indicates for the time interval between the peak of an ECG signal and a PPG signal reference point or the peak of a cycle in the SDPPG signal[10]. Unfortunately, PTT-based BP estimation alone is not accurate enough to be used for continuous cuff-less BP measurement in clinical settings[19]. However, this accuracy can be improved by the use of new BP-related features. One of the features that can increase the estimation accuracy of the regression model is Heart Rate (HR), as several studies have shown an improvement in the results after combining this feature with PTT[9]. Since the behavior of blood flow in vessels depends on various factors, PPG will also be a good signal to improve the results of BP estimation. This improvement can be made by combining PTT with several different features of PPG, one of which is the PPG Intensity Ratio (PIR).

In theory, changes in arterial diameter, △d, could be reflected by PIR throughout one cardiac cycle from systole to diastole. Moreover, there is an exponential relationship between PIR and △d that is shown by this expression[19], [20]:

$$PIR = e^{\alpha.\Delta d} \qquad (3)$$

Essentially, PIR has been defined as the maximum to minimum ratio of the amplitude of a PPG waveform. IH is the peak point of a PPG cycle or maximum amplitude , and IL is the bottommost point of a PPG cycle or minimum amplitude where α is a constant that is associated with the optical absorption coefficient in the light path. Physiologically, four variables largely influence BP, including cardiac output, arterial compliance, blood volume, and peripheral resistance. PTT could be used to evaluate arterial compliance because it has been proposed to be one of the indices of arterial stiffness[19], [21]. Moreover, there may be a relationship between cardiac output and PTT via the heart rate. Considering blood volume and peripheral resistance, changing the arterial diameter has been regarded as a main source to be evaluated by PIR that has been already illustrated. Therefore, BP changes could be directly captured by PIR and PTT employed to estimate BP[19].

The features used in this study are PTT, PIR, and HR, which are independent variables. Systolic Blood Pressure (SBP) and Diastolic Blood Pressure (DBP) are the dependent variables. After extracting these features, we developed several models based on regression on the data, but these models were found to be not sufficiently accurate because of the inconsistency and multitude of trends in the data for different features and the high dispersion of feature values. Thus, we clustered the data and developed a regression model for each cluster ,and then obtained a final estimate by averaging the outputs of these models with attention to the number of samples in each cluster.

Figure 1 describes the block diagram of the process of PB estimation with the proposed method:

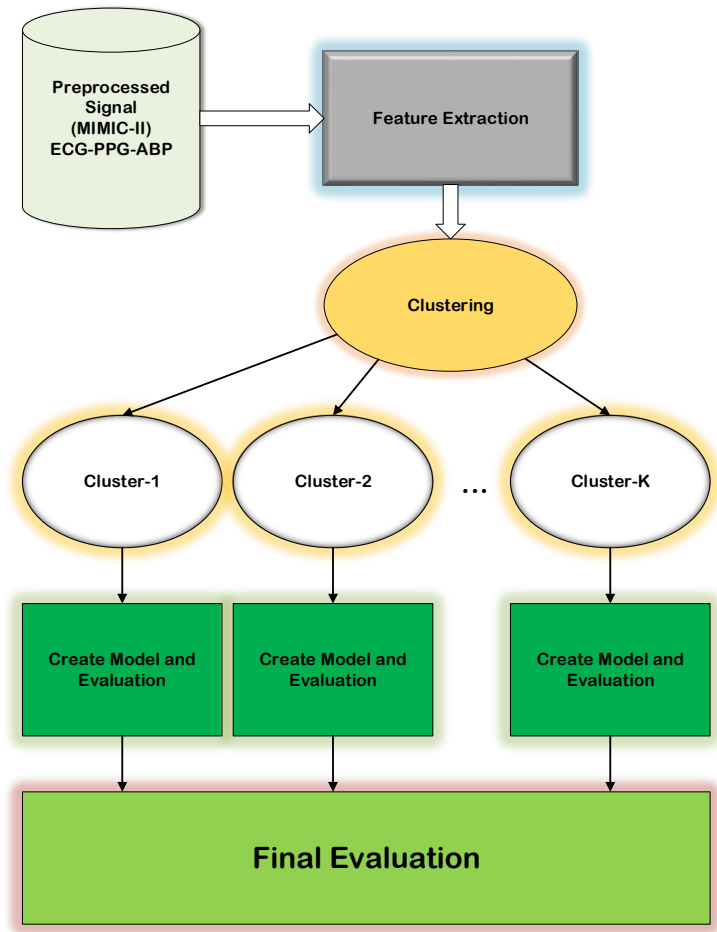

Figure 1 . Cuff-less BP estimator with clustering block diagram

Figure 2 depicts the extraction process of PIR and PTT. Now, PTT represents the time between the peak of the second derivative of PPG or SDPPG wave in the cardiac cycle and the peak of the ECG wave. As mentioned earlier, PIR has been proposed to be the ratio of minimum amplitude (IL) to maximum amplitude (IL) of a PPG signal in the cardiac cycle[10], [19], [20]. The interval between two successive QRS complexes can be used to measure the heart rate when the cardiac rhythm is regular. The heart cost is assumed on papers by divide the number of big boxes between two subsequent QRS waves by 300.

SBP and DBP may be calculated by taking the maximum and minimum values of the ABP signal in each cycle. Mathematical equations[13]:

$$SBP = max(ABP) \qquad (4)$$

$$DBP = min(ABP) \qquad (5)$$

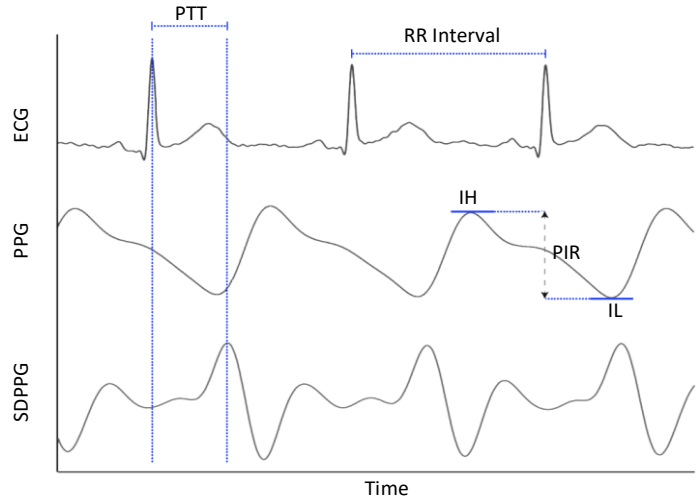

Figure 2 . Computation of the ratio of photoplethysmogram (PPG) intensity (PIR) and pulse transit time (PTT). Here, IH refers to the PPG peak intensity, SDPPG the second derivative of PPG, and finally IL valley intensity

## 2.3 Clustering and regression models

*2.3.1 Clustering.* There are several methods and algorithms for dividing a set of items into identical or highly similar clusters. The k-means algorithm is one of the simplest and most popular clustering algorithms used in data mining and unsupervised machine learning.

In multivariate clustering, it is typically needed to use multiple features of items to cluster them, which raises the question of what distance functions to use for this purpose. In any case, what is important in this clustering is the way we measure the degree of similarity or dissimilarity between data samples.

The goal of the clustering operation is to form clusters so that the distance between items in each cluster is minimal. In contrast, if the similarity of items is measured by a similarity function, the goal will be to form clusters so as to maximize the value of that function for each cluster. Given the inconsistency and multitude of trends in the data for different features and the high dispersion of feature values, we tried to first obtain clusters of data or independent variables. This was done by assessing the appropriateness of the number of clusters based on the silhouette value, and ultimately using the values of the independent variable in each cluster as the input of the regression model.

*2.3.2 Random forest regression.* Random Forest is an easy-to-use machine learning algorithm that tends to provide excellent results even without the adjustment of its meta-parameters. Thanks to its simplicity, this algorithm is one of the machine learning algorithms that are widely used for both classification and regression.

Random Forest falls in the category of supervised machine learning algorithms. As the name implies, this algorithm builds a random forest made of a group of decision trees. This is often done by the method known as bagging, the basic idea is to use a combination of learning models to reach better results. Simply put, Random Forest builds several decision trees and merges them to make more accurate and consistent predictions[22].

*2.3.3 Gradient boosting regression.* Gradient boosting is a classification and regression machine learning algorithm, which builds a prediction model using an ensemble of weak models. The goal of almost all machine learning algorithms is to minimize a defined loss function during the learning process. The constructed model needs to be updated such that the value of the loss function value approaches zero and the predicted values approach the observed values as much as possible.

The core idea of the gradient boosting algorithm is to make stronger models by combining weaker models in an iterative process.

Here, it is necessary to first describe how boosting models are created. To build boosting models, we first perform a sampling with replacement in which samples have a fixed weight in the selection probability calculations. After building a model with these samples, the samples that have produced the highest errors are returned to the sample pool and the sample selection probabilities for the next iteration of modeling are updated according to the error of each sample, which also ensures that the models properly cover the entire solution space. In the end, an ensemble of all models made through this process is created. The figure below shows the pseudocode of the boosting model construction process.

In gradient boosting regression, we first construct a regression tree model for the samples and measure the error of this model, i.e. the difference between the observed values and its predictions. We then build a new model for the data that the previous model have predicted incorrectly and recalculate the error. Next, we combine the new model with the previous one and update the ensemble. These steps are repeated until the sum of errors approaches a fixed value or the model becomes overfit[23].

*2.3.4 Deep Multilayer Perceptron.* MLP has been considered one of the supervised learning algorithms for learning a function $f(.) = R^m \to R^o$ via training on a data-set so that **m** and **o** represent the number of dimensions for input and output, respectively.

According to the target $y$ and a collection of features $X = x_1, x_2, ..., x_m$, MLP is capable of learning a nonlinear function approximator for regression and or classification. In fact, there is a difference between it and logistic regression because one or more nonlinear layers, known as hidden layers, may exist between the output and input layers. In addition, the leftmost layer that is also called input layer contains a set of neurons $\{x_i | x_1, x_2, ..., x_m\}$ implying the input features. All neuron in the hidden layer transform values from the , previous layer with a weighted linear summation $\omega_1 x_1 + \omega_2 x_2 + \cdots \omega_m x_m$ and then a nonlinear activation $g(.): R \to R$ such as the hyperbolic tan function.

*2.4 Model and Results evaluation*

*2.4.1 MAE and RMSE.* In this study, the modeling results are assessed in terms of Root Mean Square Error (RMSE) and Mean Absolute Error (MAE). Provided in the following is a description of these model evaluation criteria. The root mean square error (RMSE) quantifies how far the model's or statistical estimator's predicted values differ from the observed values. RMSE is an excellent measure for evaluating the prediction error of a model for a given dataset. This metric is basically the standard deviation of the difference between expected and observed values, as shown below:

$$RMSE = \sqrt{\frac{1}{n} \sum_{j=1}^{n} (y - \hat{y})^2} \qquad (6)$$

As many have pointed out, because of using the square root of the mean square error, RMSE is not as biased as other measures and is very suitable for medical and bioinformatics problems that are solved by regression. The other error measure used in this study is MAE. MAE measures the difference between predicted and observed values without considering the direction of this difference. Therefore, what is important for MAE is the magnitude of error in estimations not whether they have been overestimates or underestimates. In statistical discussions, this measure is sometimes referred to as L1 Loss.

Mathematically, MAE is the average absolute difference between predicted and observed value as formulated below:

$$MAE = \frac{\sum |y_i - \hat{y}_i|}{n} \tag{7}$$

*2.4.2 BHS and AAMI.* Many studies in the field of BP estimation use the protocol developed by the British Hypertension Society (BHS) for the evaluation of BP measuring devices and methods as the benchmark of their accuracy assessments. In this protocol, accuracy evaluations are performed based on the absolute error of measurements. More specifically, this protocol grades the methods and devices based on the ratio of the number of readings with an error of less than 5 mmHg, 10 mmHg, and 15 mmHg to the total number of readings.

Another standard for evaluating BP measuring devices and methods is the AAMI standard. In this standard, a device or method is approvable only if the mean error and standard deviation of readings are less than 5 mmHg and 8 mmHg, respectively. In this study, the accuracy of SBP and DBP estimates is evaluated using BHS and AAMI standards.

# 3. Results

First, the data and the extracted features were visualized and the correlation between the features was measured. A very important section of creation regression model is the preparation of data, which in this study involved a scaling operation. This phase is very important because it affects how much time it takes to construct the regression model and the length of the convergence process. Next, we developed several machine learning regression models , including random forest regression, gradient boosting regression, and multilayer perceptron regression and evaluated the model outputs by different criteria. The next step was to implement the main approach of the study, that is, to cluster the extracted data or features and variables while using the silhouette value to determine the best number of clusters and then develop a model for each cluster with the regression algorithms mentioned above. the model outputs were compiled by weighted averaging ,and the final results were compared in terms of different measures to identify the best regression model.

Figure 3 shows the histogram and scatter diagram of PTT, PIR, BPMIN, and BPMAX. We used the scatter diagram to create a graphical representation of the relationship between independent and dependent features and we plotted a density diagram to gain an overview of the distribution of values for each feature.

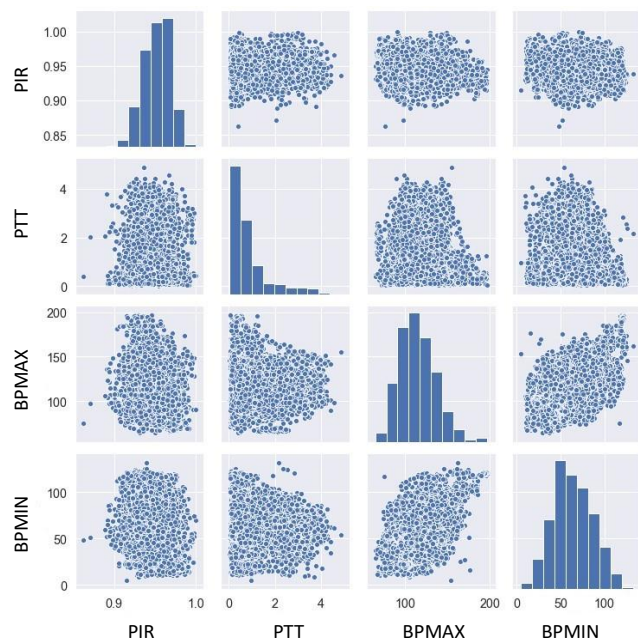

Figure 3 . histogram and scatter diagram of the features PTT, PIR, BPMIN, and BPMAX.

The next step was to obtain and examine the results of the machine learning regression models described in the previous sections. In this step, the machine learning models were developed with the features PTT, PIR, HR as independent variables (input) and bpmin and bpmax as dependent variables (output). First, we developed the model by regression on the entire data using random forest regression, gradient boosting regression, and multilayer perceptron regression. The results of this process are presented in Table 1. It should be noted that in all steps, the regression models were evaluated in terms of RMSE and MAE.

Table 1. COMPARISON OF THE PERFORMANCE WITHOUT CLUSTERING
AND REGRESSION ALGORITHMS

|  | Systolic Blood Pressure (mmHg) | | | Diastolic Blood Pressure (mmHg) | | |
|---|---|---|---|---|---|---|
| **Learner/Performance** | MAE | RMSE | r | MAE | RMSE | r |
| **Random Forest Regression** | 7.426 | 12.250 | 0.65 | 7.410 | 12.110 | 0.68 |
| **Gradient Boosting Regression** | 6.367 | 10.395 | 0.67 | 6.276 | 10.221 | 0.71 |
| **MultiLayer Perceptron Regression** | 9.422 | 14.120 | 0.59 | 9.323 | 14.099 | 0.64 |

We utilized the k-means method to cluster the data using the Silhouette criteria to identify the optimal number of clusters, given the inconsistency and variety of trends in the data for different features. Figure 4 shows the optimal number of clusters for the clustering algorithm according to the Silhouette criterion.

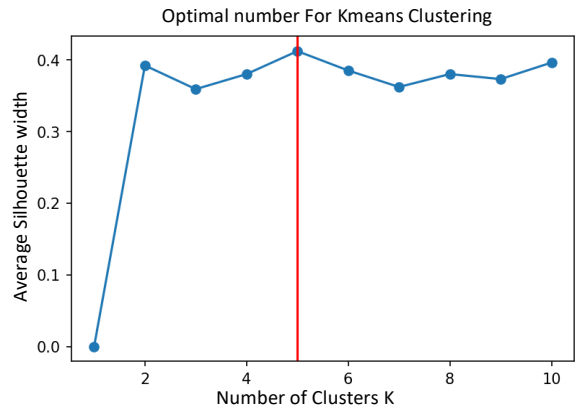

Figure 4. Optimal number of clustering with Silhouette criteria

Figure 5 shows the cohesion and dispersion of data in one of the clusters extracted from the data PIR, PTT, SBP and DBP.
Next, we used random forest regression, gradient boosting regression, and multilayer perceptron regression algorithms to develop a separate model for each cluster.
The model error for each cluster was then determined in terms of RMSE and MAE and target-estimation correlation coefficient(r) for Gradient boosting regression. Finally, the total error of the model and correlation coefficient for all clusters was determined by Weighted arithmetic mean. Finally, the error rate for the whole data and the total error rate are also provided. The results of the proposed clustering-based approach are presented in Table 2.

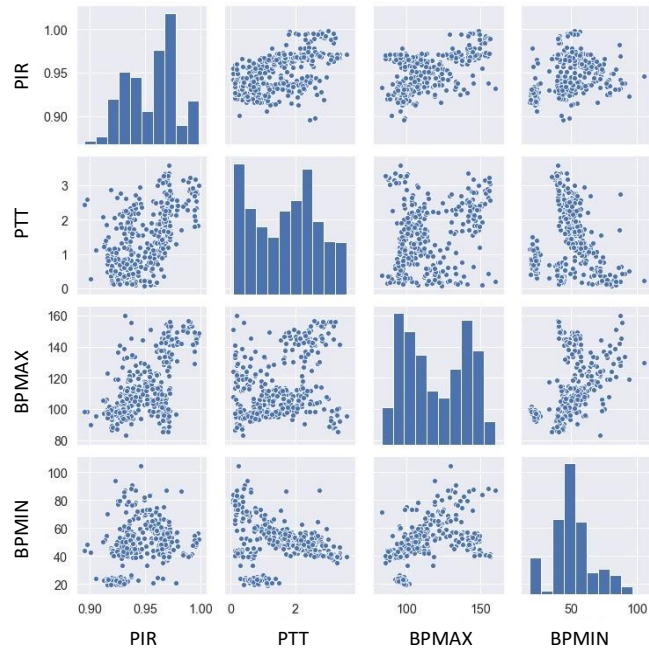

Figure 5 . histogram and scatter diagram of the features PTT, PIR, BPMIN,
and BPMAX in one of the clusters

Table 2 . COMPARISON OF THE PERFORMANCE WITH CLUSTERING AND REGRESSION ALGORITHMS

| Learner/Performance | Cluster | Count of data per cluster | Systolic Blood Pressure (mmHg) | | | Diastolic Blood Pressure (mmHg) | | |
|---|---|---|---|---|---|---|---|---|
| | | | MAE | RMSE | r | MAE | RMSE | r |
| **Random Forest Regression** | Cluster1 | 6282 | 3.407 | 5.830 | - | 3.250 | 5.698 | - |
| | Cluster2 | 6276 | 3.468 | 5.724 | - | 3.038 | 5.136 | - |
| | Cluster3 | 3355 | 3.521 | 5.586 | - | 2.813 | 5.408 | - |
| | Cluster4 | 8300 | 3.396 | 5.500 | - | 2.870 | 4.852 | - |
| | Cluster5 | 2390 | 2.434 | 4.567 | - | 2.677 | 4.879 | - |
| | Total | **26603** | **3.344** | **5.557** | **-** | **2.974** | **5.191** | **-** |
| **Gradient Boosting Regression** | Cluster1 | 6282 | 2.644 | 5.841 | 0.96 | 2.486 | 5.648 | 0.98 |
| | Cluster2 | 6276 | 2.781 | 5.694 | 0.93 | 2.468 | 5.232 | 0.96 |
| | Cluster3 | 3355 | 2.533 | 6.123 | 0.76 | 2.003 | 5.491 | 0.80 |
| | Cluster4 | 8300 | 2.610 | 5.522 | 0.85 | 2.161 | 4.675 | 0.95 |
| | Cluster5 | 2390 | 1.643 | 4.709 | 0.85 | 1.504 | 4.467 | 0.95 |
| | Total | **26603** | **2.561** | **5.635** | **0.88** | **2.231** | **5.012** | **0.94** |
| **MultiLayer Perceptron Regression** | Cluster1 | 6282 | 5.230 | 8.244 | - | 4.896 | 7.262 | - |
| | Cluster2 | 6276 | 5.340 | 8.754 | - | 5.263 | 8.956 | - |
| | Cluster3 | 3355 | 6.235 | 9.523 | - | 6.094 | 8.852 | - |
| | Cluster4 | 8300 | 7.261 | 11.920 | - | 6.288 | 9.003 | - |
| | Cluster5 | 2390 | 4.326 | 8.156 | - | 4.160 | 6.875 | - |
| | Total | **26603** | **5.937** | **9.664** | **-** | **5.501** | **8.370** | **-** |

Figure 6 shows the Bland-Altman plot of each cluster for SBP and DBP estimations. As the figure shows, most errors are in the range of 8 mmHg for DBP and 12 mmHg for SBP. However, there are also some outliers in these plots, which are more frequent in the one for SBP estimation.

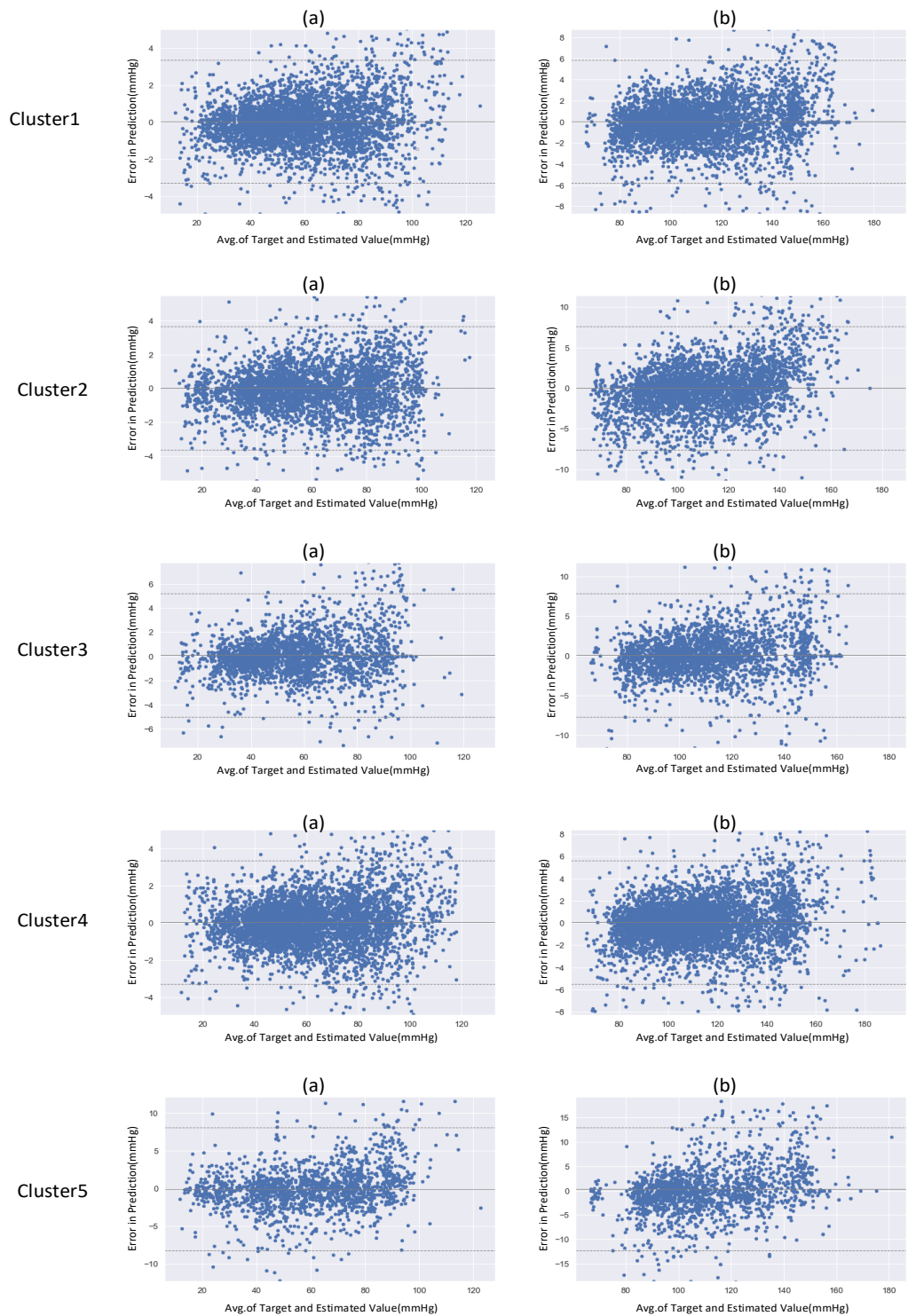

Figure 6 . Bland-Altman Plot for DBP (a) and SBP (b) estimations in each cluster

Figure 7 shows the correlation plots of DBP and SBP estimatation for our suggested technique vs reference BP. The overall calculated DBP and SBP had a correlation value of 0.94 and 0.88, respectively. which obtained by weighted averaging of correlation coefficient in each cluster.

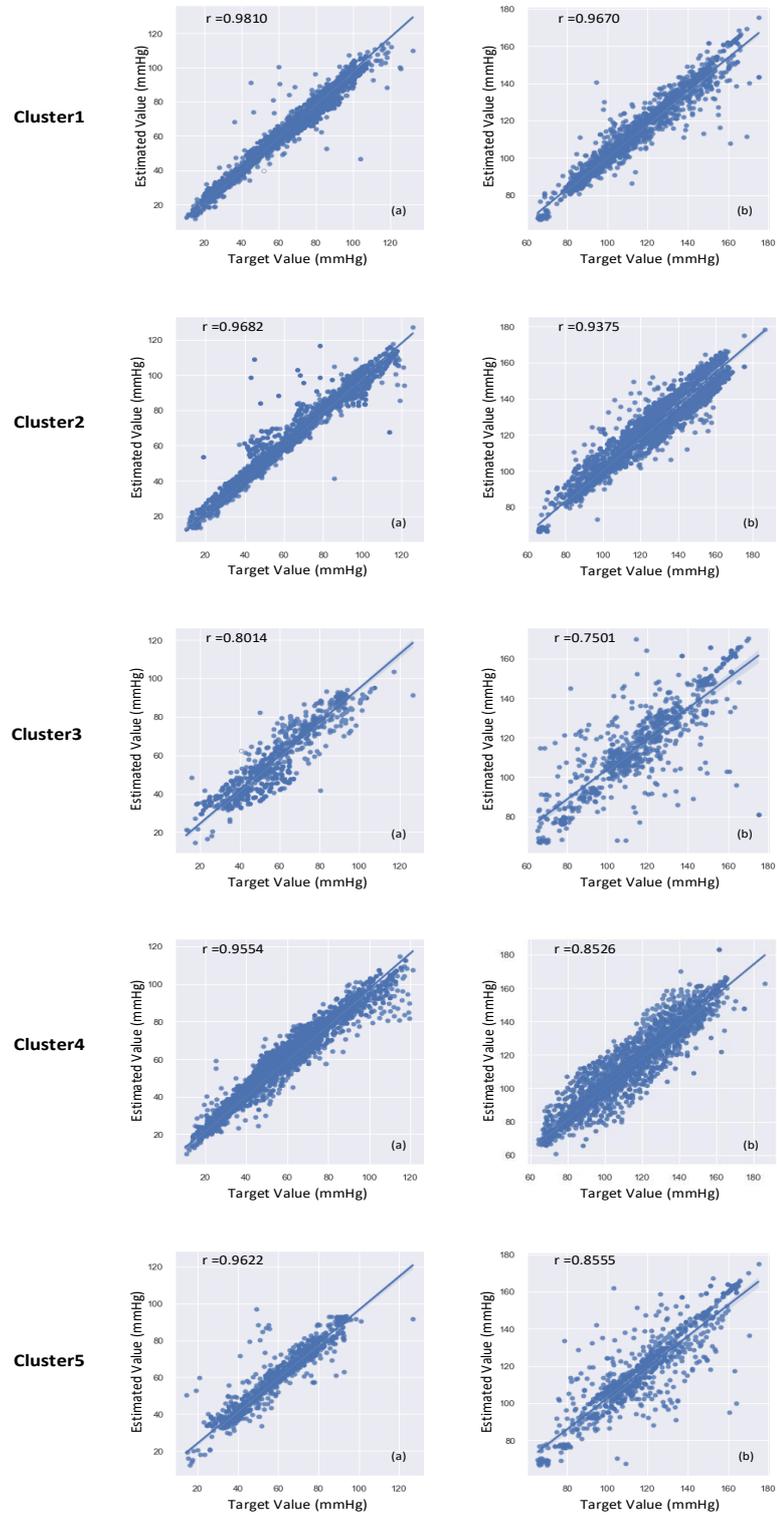

Figure 7 .Correlation Plot for DBP (a) and SBP (b) estimations in each cluster

After applying clustering on the sample of features extracted from ECG, PPG and ABP signals, we investigated the raw signal corresponding to each data sample in each cluster. Evidence showed that the ECG, PPG and ABP signals corresponding to each data sample in each cluster are very similar in appearance and signal shape, which can be used to study the physiological characteristics of patients.

Figure 8 shows the ECG, PPG and ABP signals of two different patients in cluster1 and cluster2 which are very similar to same cluster samples and very different from other cluster samples:

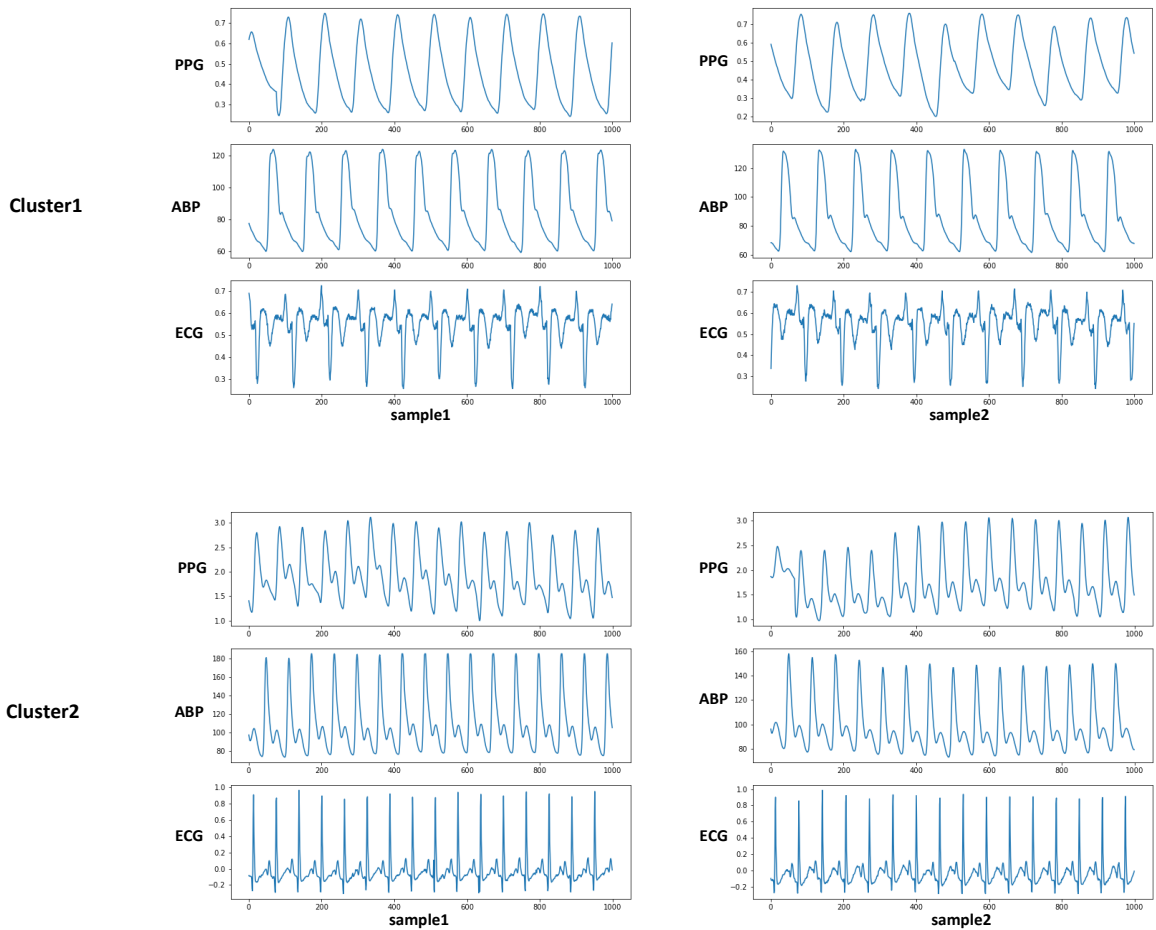

Figure 8 . PPG , ABP and ECG signals related to two patients in cluster1 and cluster2

The results of the accuracy evaluation of the proposed method based on the BHS standard are presented in Table 3. According to this standard, the proposed method will be of Grade A in DBP estimation and SBP estimation.

Table 3 . BHS Standard assessment for all clusters with weighted averaging

|  |  | Cumulative Error Percentage | | |
|---|---|---|---|---|
|  |  | ≤ 5 mmHg | ≤ 10 mmHg | ≤ 15 mmHg |
| Our Results | **DBP** | **73.05 %** | **90.12 %** | **97.34 %** |
|  | **SBP** | **65.59 %** | **86.54 %** | **96.32 %** |
| BHS | grade A | 60 % | 85 % | 95 % |
|  | grade B | 50 % | 75 % | 90 % |
|  | grade C | 40 % | 65 % | 85 % |

Table 4 shows the results of the accuracy evaluation of the proposed method based on the AAMI standard. According to this standard, the method produces acceptably accurate estimates for DBP and SBP.

Table 4 . AAMI Standard assessment for all clusters with weighted averaging

|  |  | ME (mmHg) | STD (mmHg) | Number of Subjects |
|---|---|---|---|---|
| Our Results | **DBP** | **2.811** | **5.596** | **942** |
|  | **SBP** | **3.987** | **5.715** | **942** |
| AAMI Standard |  | ≤ 5 | ≤ 8 | ≥ 85 |

# 4.Discussion

It should be noted that while a large number of studies have been conducted in the field of BP estimation, many of these studies have used their own datasets, the majority of which are not publically available due to confidentiality and privacy considerations. Therefore, we cannot compare our results with all of the previous studies. In this section, we first compare our results with the results of studies that have used MIMIC/PhysioNet datasets, and then make some comparisons with studies that have used their own datasets. A noteworthy point regarding the MIMIC-II dataset is that it comprises readings from ICU patients, who tend to be older and under medication[9]. Another important point regarding MIMIC-II is the lack of physiological data (e.g. age, height, weight, etc.), which can affect the accuracy of the extracted features and the model. While we could potentially use these data to include physiological and biological parameters in the clusters and examine their effects on the estimation accuracy, unfortunately, this could not be done with MIMIC-II. Because of using MIMIC-II, in this study, we only had access to ECG, PPG, and ABP signals, and therefore our feature extraction was limited to these signals. Thus, using a richer dataset containing other signals such as SCG and BCG in addition to ECG, PPG and ABP may be able to improve the accuracy of the extracted features and the resulting model[24].

The studies that have used publicly available MIMIC datasets include [9] and [18], where PAT, HR, AI, LASI, and IPA features were extracted from ECG and PPG signals and then the Adaboost algorithm was used to develop BP estimation models based on these features. Our method outperforms the models of [9] and [18] in terms of MAE and r as well as BHS and AAMI standards. Our results are also better than those reported in [11], where an estimation model was developed by multi-sample regression based on 35 features extracted from the same ECG and PPG signals.

In [13], Ibtehaz et al. developed their estimation model with the CNN algorithm using only the PPG signal. Our method also performs better than this model in terms of MAE and BHS and AAMI standards. Our results are also more accurate than the results of [12], where Kurylyak et al. used 21 features extracted from ECG and PPG signals of MIMIC-II and an ANN algorithm to develop their model. The same can also be said for [25] and [26], where estimation was performed using the features extracted from the MIMIC dataset of the PhysioNet website.

We also compared our method with some of the methods that have used their own datasets, which are listed in Table 5. In [17], Chen et al. created their own dataset by compiling the data of 98 subjects and developed their model using the multiple regression method based on features like PTT. Our method showed better performance in estimating SBP and DBP than this model. Our results are also better than the results of [26], where Radha et al. used a dataset consisting of the data collected from 106 healthy individuals with Random Forest and Dense Network, and also the results of [27], where Esmaili et al. used a dataset compiled from the data of 32 subjects with a calibration step. The results of the present work are also more accurate than those of [28], [29] and other listed works that have used their own datasets.

Table 5 . COMPARISON WITH OTHER WORKS

| Studies | Method | Machine learning comparison (DBP) | | | Machine learning comparison (SBP) | | |
|---|---|---|---|---|---|---|---|
| | | MAE | RMSE | r | MAE | RMSE | r |
| Our Proposed Method | Clustering & Gradient Boosting Regression | **2.23** | **5.01** | **0.94** | **2.56** | **5.63** | **0.88** |
| Our Proposed Method | Gradient Boosting Regression without Clustering | **6.27** | **10.22** | **0.71** | **6.36** | **10.39** | **0.67** |
| [16] | SVM | 6.34 | - | - | 12.38 | - | - |
| [9] | Adaboosting | 5.35 | - | 0.48 | 11.17 | - | 0.59 |
| [17] | MLR | 2.82 | - | 0.97 | 2.83 | - | 0.96 |
| [26] | Lstm & perceptron | - | 6.49 | | | 7.86 | |
| [11] | Multi sensor features | 4.54 | - | 0.90 | 6.13 | - | 0.84 |
| [13] | PPG+CNN-Regression | 3.45 | - | 0.89 | 5.73 | - | 0.93 |
| [19] | PTT+PIR+ Nonlinear Regression | 3.18 | - | 0.88 | 4.09 | - | 0.91 |
| [28] | - | 3.27 | - | 0.87 | 4.46 | - | 0.93 |
| [27] | (PPG+ECG) | 4.44 | - | 0.84 | 4.71 | - | 0.89 |
| [30] | SVM | 3.36 | - | 0.82 | 11.86 | - | 0.69 |
| [29] | MLP | 4.96 | - | 0.70 | 5.46 | - | 0.87 |
| [31] | ECG: wrist & foot PPG: finger | 4.4 | - | - | 6.0 | - | - |
| [32] | ANN with 15 hidden neurons | Not mentioned | - | - | 3.03 | - | - |
| [33] | PTT & PIR,Regression-MARS | 4.86 | - | 0.93 | 7.83 | - | 0.95 |
| [25] | AutoML(TPOT) | 4.19 | - | - | 6.52 | - | - |

| [12] | ANN | 2.21± 2.09 | - | - | 3.80±3.46 | - | - |
| [26] | DNN | 6.88 | - | - | 9.43 | - | - |
| [34] | Res-LSTM | 4.61 | - | 0.74 | 7.10 | - | 0.96 |
| [35] | LSTM-base autoencoder | 4.05 | - | - | 2.41 | - | - |

## 5. Conclusion and future works

This study developed a new clustering-based algorithm to improve the accuracy of the blood pressure estimation, which uses the K-means algorithm for Clustering extracted features and random forest regression algorithm, gradient boosting regression algorithm , and multilayer perceptron regression algorithm to estimate Systolic Blood Pressure (SBP) and Diastolic Blood Pressure (DBP) in each cluster. The results showed that, according to high dispersion and the multitude of trends in the data and extracted features, the clustering algorithm can increase the prediction accuracy for each model. Overall, it can be concluded that since previous works have chosen not to deal with high dispersion and multitude of trends in the data before developing their learning models, it is indeed possible to reach considerably better prediction results by applying a clustering algorithm to the extracted data and then building a separate model for each cluster. In future works, we hope to develop a method for real-time feature extraction and sample clustering and ultimately create a real-time procedure for receiving vital signals such as ECG and PPG from thousands of people, performing feature extraction and signal processing, clustering the data, and producing BP estimates with the least possible delay and the highest possible accuracy; a task that will require using Big Data related platforms, tools, and algorithms.

## Data Availability

The data for this study originated from PhysioNet and the well-known MIMIC-II database, however a preprocessed dataset from the MIMIC-II database is available at https://www.kaggle.com/mkachuee/BloodPressureDataset , which we utilized.

## Conflicts of Interest

The authors declare that they have no conflicts of interest.